\documentclass[aps,prd,preprint,showpacs,superscriptaddress]{revtex4-1}

\usepackage{graphics}

\usepackage{enumerate}
\usepackage{amsmath,amsthm,amssymb}
\usepackage{slashed}
\usepackage{array}
\usepackage{latexsym}

\usepackage{graphicx}

\usepackage{pstricks}

\usepackage{epsfig}

\begin{document}

\title{Texture Zeros for the Standard Model Quark Mass Matrices}

\author{William A. Ponce}
\affiliation{Instituto de F\'\i sica, Universidad de Antioquia,
A.A. 1226, Medell\'\i n, Colombia.}

\author{Richard H. Benavides}
\affiliation{Instituto de F\'\i sica, Universidad de Antioquia,
A.A. 1226, Medell\'\i n, Colombia.}
\affiliation{Instituto Tecnol\'ogico Metropolitano, Facultad de Ciencias, Medell\'in, Colombia}


\date{\today}

\begin{abstract}
A way of counting free parameters in the quark mass matrices of the standard model, including the constraints 
coming from weak basis transformations, is presented; this allow to understand the exact physical meaning of 
the parallel and non-parallel texture zeros which appear in some ``ans\"{a}tz'' of the $3\times 3$ quark mass 
matrices, including the CP violation phenomena in the analysis, it is shown why the six texture zeros are ruled out. 
Finally, a five texture zeros  ``ans\"{a}tze''which properly copes with all experimental constrains, including the 
angles of the unitary triangle, is presented.
\end{abstract}

\maketitle

\section{Introduction}
\label{intro}
Although the gauge boson sector of the Standard Model (SM) with the $SU(3)_c\otimes SU(2)_L\otimes U(1)_Y$  local symmetry has been very successful~\cite{SM}, its Yukawa sector is still poorly understood. Questions related with this sector as for example: the total number of families in nature, the hierarchy of charged fermion mass spectra, the smallness of neutrino masses, the quark and lepton mixing angles, and the origin of the CP violation, remain until today as open problems in theoretical particle physics.

To date, several approaches have been suggested in the literature in order to understand the phenomenology of the Yukawa sector; among them: radiative mechanisms~\cite{wein}, horizontal symmetries; discrete~\cite{Dis}, and continuous, global, and local gauge symmetries~\cite{Con}, which may or may not include the so-called Froggat and Nielsen mechanism~\cite{Fro}. Phenomenologically, a common approach is to search for simple textures of quark mass matrices that can predict self-consistent and experimentally favored relations between quark masses and flavor mixing parameters~\cite{Fri,Ram}.

In the SM and after the local gauge symmetry has been spontaneously broken, the quark mass terms are given by
\begin{equation}\label{lama}
-{\cal L}_m=\bar U_{0L}M_u U_{0R}+\bar D_{0L}M_dD_{0R}+h.c,
\end{equation}
where $\bar U_{0L}=(\bar u_0,\bar c_0,\bar t_0)_L,\;\; \bar D_{0L}=(\bar d_0,\bar s_0,\bar b_0)_L,\;\;
U_{0R}^T=(u_0,c_0,t_0)_R,\;\;D_{0R}^T=(d_0,s_0,b_0)_R,$ (where the upper $T$ stands for transpose, and the down zero stands for weak basis quark states). The matrices  $M_u$ and $M_d$ in (\ref{lama}) are $3\times 3$ complex mass matrices. In the most general case they contain  36 free parameters. In the context of the SM, such a large number of parameters can be drastically cut by making use of the polar theorem of matrix algebra, by which, one can always decompose a complex matrix as the product of an Hermitian times a unitary matrix. Since for the SM the unitary matrix can be absorbed in a redefinition of the right handed quark components, this immediately brings the number of free parameters from 36 down to 18 (the other eighteen parameters can be hidden in the right-handed quark components in the context of the SM and some of its extensions, but not in its left-right symmetric extensions).

So, as far as the SM is concerned we may treat, without loss of generality, $M_u$ and $M_d$ as two Hermitian quark mass matrices, with 18 parameters in total, out of which 6 are phases. Since 5 of those phases can be absorbed in a redefinition of the quark fields~\cite{koma}, the total number of free parameters we may play with in $M_u$ and $M_d$ are 12 real parameters and one phase; this last one used to explain the CP violation phenomena.

In what follows we are going to present numeric and analytic results for three sets of SM quark mass matrices, two of them containing six texture zeros, taking special care to accommodate the latest experimental data available~\cite{pdg}, including the CP violation phenomena. In our approach we make use of the so called weak basis transformations technique, which change the given quark mass matrices and its related mixing matrix into equivalent ones~\cite{Bra1}, applying such a technique to place texture zeros at the (1,1) entries of both quark mass matrices, resembling a kind of see-saw mechanism for the first family of quarks.

This paper is organized as follows: in Sec.~\ref{sec:sec2} some features of the SM mixing matrix are presented; in Sec.~\ref{sec:sec3} the concept of weak basis transformation is review; in Sec.~\ref{sec:sec4} we present our study of the parallel six texture zeros Fritzsch ``ans\"atze'' and in Sec.~\ref{sec:sec5} we analyze a non parallel six texture zeros ``anz\"atze''. Sec.~\ref{sec:sec6} is devoted to the study of a five texture zeros ``anz\"atze'' which works properly. Our conclusions are presented in Sec.~\ref{sec:sec7}. The values of the quark masses used in the numerical studies are quoted at the end of the paper in an appendix.

\section{\label{sec:sec2}The SM Mixing Matrix}

In the SM and for the six flavor case, the Baryon charged weak current is given by 
\begin{equation}\label{weakcurr}
J_{\mu L}^-=\bar U_{0L}\gamma_\mu D_{0L}=\bar U_L\gamma_\mu V_{CKM}D_L,
\end{equation}
where  $V_{CKM}=U_u U_d^\dag$ is the Cabibbo-Kobayashi-Maskawa (CKM) mixing matrix, with $U_u$ and $U_d$ the unitary matrices which diagonalize the Hermitian $M_uM_u^\dagger$ and $M_dM_d^\dagger$ square mass matrices respectively, and $\bar U_{L}=(\bar u,\bar c,\bar t)_L$ and $D_{L}^T=(d,s,b)_L$ stand for the quark field mass eigenstates.

$V_{CKM}$ is a $3\times 3$ unitary matrix, its form is not unique, but the permutation freedom between the three generations can be  removed by ordering the families such that $(u_1,u_2,u_3)\rightarrow(u,c,t)$ and $(d_1,d_2,d_3)\rightarrow(d,s,b)$. The complex elements of $V_{CKM}$ are thus commonly written as 
\begin{equation}\label{vdef}
 V_{CKM}=\left(\begin{array}{ccc}
V_{ud} & V_{us} & V_{ub} \\
V_{cd} & V_{cs} & V_{cb} \\ 
V_{td} & V_{ts} & V_{tb} 
               \end{array}\right).
\end{equation}
The unitary of the CKM mixing matrix leads to relations among the rows and columns of $V_{CKM}$, in particular we have for the columns: 
\begin{align}\label{ts}
V_{ud}V_{us}^*+V_{cd}V_{cs}^*+V_{td}V_{ts}^*&=0, \\ \label{tc} 
V_{us}V_{ub}^*+V_{cs}V_{cb}^*+V_{ts}V_{tb}^*&=0, \\ \label{tb}
V_{ud}V_{ub}^*+V_{cd}V_{cb}^*+V_{td}V_{tb}^*&=0. \\ \nonumber
\end{align}
Each of these three relations requires the sum of three complex quantities to vanish and so can be geometrical represented in the complex plane as a triangle. These are the unitary triangles~\cite{nir}, though the term ``unitary triangle is usually reserved for the relation (\ref{tb}) only.

The three angles of the unitary triangle  represented by (\ref{tb}), which are physical quantities and can be independently measured by CP asymmetries in B decays. are defined as follows~\cite{nir}:
\begin{eqnarray}\label{alpha}
\alpha&\equiv&{\rm arg}\left[-\frac{V_{td}V^*_{tb}}{V_{ud}V_{ub}^*}\right]\\ \label{beta}
\beta&\equiv&{\rm arg}\left[-\frac{V_{cd}V^*_{cb}}{V_{td}V_{tb}^*}\right]\\ \label{gamma}
\gamma&\equiv&{\rm arg}\left[-\frac{V_{ud}V^*_{ub}}{V_{cd}V_{cb}^*}\right],\\ \nonumber
\end{eqnarray}
 The experimental findings at the B factories, fitted to close the triangle, are~\cite{ckmf,babar}
\begin{equation}\label{utex}
(\alpha,\;\beta,\;\gamma)_{exp}^{fit}=(91.0\pm 7.2,\;\;\;21.8\pm 2.8,\;\;\;67.2\pm 9.1),
\end{equation}
with an accuracy in the measurement of $\sin 2\beta$ no less than 20\%~\cite{sony}.

The SM mixing matrix $V_{CKM}$ has three mixing angles $\theta_{ij},\;\; i<j,\;\; i,j=1,2,3,$ the mixing angles between the $i^{th}$ and $j^{th}$ families, and only one CP violating phase~\cite{koma}. It has been parametrized in the literature in several different ways, but the most important fact related with this matrix is that most of its entries have been measured with high accuracy, with the following experimental bounds~\cite{pdg,ckmf}:
\begin{equation}\label{maexp}
 V^{(exp)}=
\left(\begin{array}{ccc}
0.970\leq |V_{ud}|\leq 0.976 & 0.222\leq |V_{us}|\leq 0.226 & 0.003\leq |V_{ub}|\leq0.004\\
0.217\leq |V_{cd}|\leq 0.237 & 0.960\leq |V_{cs}|\leq 0.990   & 0.039 \leq |V_{cb}|\leq 0.041\\
0.008\leq |V_{td}|\leq 0.009 & 0.038\leq |V_{ts}|\leq 0.042 & 0.999\leq |V_{tb}| < 1.000
\end{array}\right),
\end{equation}
where the experimental numbers quoted above at 95\% C.L. are restricted to fit the unitary conditions of $V_{CKM}$ due to the fact that we are going to confront them with quark mass matrices which must fit the SM constraints.


\section{\label{sec:sec3}Weak Basis Transformations}
In the context of the SM, the most general weak basis (WB) transformation that leaves the two $3\times 3$ quark mass matrices Hermitian, and do not alter the physics implicit in the weak currents, is a unitary transformation acting simultaneously in the up and down quark mass matrices~\cite{Bra1}. That is
\begin{equation}\label{2aa}
\begin{split}
M_u&\longrightarrow M_u^R=U M_u U^\dag,\\
M_d&\longrightarrow M_d^R=U M_d U^\dag,
\end{split}
\end{equation}
where $U$ is an arbitrary unitary matrix. We say then that the two representations $(M_{u},M_{d})$ and $(M_{u}^R,M_{d}^R)$ are equivalent in the sense that they are related to the same $V_{CKM}$ mixing matrix. This kind of transformation plays an important role in the study of the so-called flavor problem.

That a WB transformation does not change the mixing matrix $V_{CKM}$ can be seen from its definition. After the WB transformation implicit in Eq. (\ref{2aa}) is done, the new mixing matrix is such that \[V_{CKM}^R =U_u^R U_d^{R\dag}=U_uU^\dag UU_d^\dag=U_uU_d^\dagger=V_{CKM},\]
with $U_u^R$ and $U_d^R$ the unitary matrices which diagonalize the Hermitian $M_u^RM_u^{R\dagger}$ and $M_d^RM_d^{R\dagger}$ square mass matrices respectively.

In the last paper of Ref.~\cite{Bra1} it has been shown that it is always possible to perform a weak basis transformation such that $(M_u^R)_{11}=(M_d^R)_{11}=(M_u^R)_{13}=(M_u^R)_{31}=0$; or equivalently $(M_u^R)_{11}=(M_d^R)_{11}=(M_d^R)_{13}=(M_d^R)_{31}=0$ (texture zeros related somehow to the mass hierarchy $m_u<m_c<m_t$ and $m_d<m_s<m_b$). The meaning of this is that it is always possible to have mass matrices with 3 texture zeros which do not have any physical meaning. With 3 texture zeros the number of free parameters in $M_u^R$ and $M_d^R$ reduces from 12 to 9 real plus one phase, just enough to fit the measured values for the 6 quark masses, the 3 mixing angles, and the CP violation phenomena. Any extra texture zero can only be a physical assumption and should imply a relationship between the quark masses and the parameters of the mixing matrix. We will elaborate on this argument in what follows.


\section{\label{sec:sec4}Parallel six texture zeros}
To explain in the context of the SM the quark mass spectrum and its mixing matrix, Harald Fritzsch proposed some time ago, the existence of texture zeros in the quark mass matrices~\cite{Fri}. Let us study his ``ans\"atze":

According to ``Fritzsch" original hypothesis, the up and down quark mass matrices assume a similar texture, named in the literature as the nearest neighbor interaction form. After redefining the right-handed quark fields, the up and down quark mass matrices assume the following particular Hermitian form

\begin{equation}\label{text}
M_q^{(6)} =\left(\begin{array}{ccc} 
0 & a_qe^{i\alpha_q} & 0 \\ 
a_qe^{-i\alpha_q}_q & 0 & b_qe^{i\beta_q}  \\
0 & b_qe^{-i\beta_q}  & c_q
\end{array}\right),
\end{equation}
where $q$ stands for $u$ and $d$. For this particular ``ans\"atze" there are six real parameters and four different phases. According to our way of counting the number of parameters, with the six  real parameters we must explain the values for the 6 quark masses and the 3 mixing angles; as we are going to show, this ``ans\"atze"  must predict three relationships between the quark masses and the parameters of the mixing matrix. Notice by the way that the invariant det$M_q^{(6)}=-c_qa_q^2<0$ for $c_q>0$.

To start, let us see that the four phases in (\ref{text}) can be absorbed by redefining new quark fields as follows:

\begin{equation}\label{upph}
\left(\begin{array}{c} u_0 \\c_0\\t_0 \end{array} \right) =
\left(\begin{array}{ccc}
1 & 0 & 0 \\ 
0 & e^{-i\alpha_u} & 0   \\
0 & 0 & e^{-i(\alpha_u+\beta_u)}
\end{array}\right)\left(\begin{array}{c} u_0^\prime \\c_0^\prime \\t_0^\prime  \end{array} \right)=U_u^\dagger\left(\begin{array}{c} 
u_0^\prime \\c_0^\prime \\t_0^\prime \end{array} \right),
\end{equation}
with the corresponding redefinition for the down sector with the replacement $\alpha_u,\beta_u\rightarrow\alpha_d,\beta_d$.

So, in the primed basis, the algebra reduces to diagonalize the two real symmetric mass matrices 
\begin{equation}\label{texp}
M_q^{(6)\prime} =\left(\begin{array}{ccc} 
0 & a_q & 0 \\ 
a_q & 0 & b_q  \\
0 & b_q  & c_q
\end{array}\right),
\end{equation}
job that can be achieved by using orthogonal transformations $O_q^{(f6)}$ (instead of the bi-unitary transformations required in the most general case). So, we have
\begin{equation}\label{mdiag}
M_q^{{\rm diag}}=O_q^{(f6)T}M_q^{(6)\prime} O_q^{(f6)}={\rm diag}(m_1,-m_2,m_3),
\end{equation}
where the sub-indices 1,2,3 in the diagonal forms refer respectively to the masses for the quarks $u,c$ and $t$ for the up sector, as well as $d,s$ and $b$ for the down sector.

Using the invariants tr$[M_q^{(6)\prime}]$, tr$[(M_q^{(6)\prime})^2]$, and det$[M_q^{(6)\prime}]$, we may write:
\begin{equation}\label{cq}
\begin{split}
c_q&=m_1-m_2+m_3\\ 
a_q^2&=\frac{m_1m_2m_3}{m_1-m_2+m_3}\\ 
b_q^2&=\frac{(m_3-m_2)(m_3+m_1)(m_2-m_1)}{m_1-m_2+m_3}.
\end{split}
\end{equation}
The exact diagonalizing transformation $O_q^{(f6)}$ for this particular ``ans\"atze" is expressed as
{\small
\begin{equation}\label{rotf6}
O_q^{(f6)}=\left(\begin{array}{ccc} \pm\sqrt{\frac{m_2m_3(m_3-m_2)}{(m_3-m_1)(m_2+m_1)(m_1-m_2+m_3)}} & \pm\sqrt{\frac{m_1(m_3-m_2)}{(m_3-m_1)(m_1+m_2)}} &  \mp\sqrt{\frac{m_1(m_2-m_1)(m_1+m_3)}{(m_3-m_1)(m_1+m_2)(m_1-m_2+m_3)}} \\ 
\pm\sqrt{\frac{m_1m_3(m_1+m_3)}{(m_2+m_1)(m_3+m_2)(m_1-m_2+m_3)}}  & \mp\sqrt{\frac{m_2(m_1+m_3)}{(m_2+m_3)(m_1+m_2)}} & \pm\sqrt{\frac{m_2(m_3-m_2)(m_2-m_1)}{(m_2+m_1)(m_3+m_2)(m_1-m_2+m_3)}}\\ 
\pm\sqrt{\frac{m_1m_2(m_2-m_1)}{(m_2+m_3)(m_3-m_1)(m_1-m_2+m_3)}} &
\pm\sqrt{\frac{m_3(m_2-m_1)}{(m_2+m_3)(m_3-m_1)}} & 
\pm\sqrt{\frac{m_3(m_3-m_2)(m_1+m_3)}{(m_2+m_3)(m_3-m_1)(m_1-m_2+m_3)}}\end{array}\right),
\end{equation}}
\noindent
where one has the freedom to choose two equivalent possibilities of phases (the up or down signs). 

For the up quark sector, and due to the fact that $m_t\gg m_c\gg m_u$, the orthogonal matrix (\ref{rotf6}) can be expanded as 
{\small
\begin{equation}\label{rot6ex}
O_u^{(f6)}\approx\left(\begin{array}{ccc} 
\pm (1-m_{uc}/2) & \pm\sqrt{m_{uc}}(1-m_{ct}/2-m_{uc}/2) & \mp\sqrt{m_{ut}}(1-m_{uc}+m_{ct}/2)  \\ 
\pm\sqrt{m_{uc}}(1-m_{uc}/2) & \mp(1-m_{ct}/2-m_{uc}/2) & \pm\sqrt{m_{ct}}(1-m_{uc}-m_{ct}/2)\\ 
\pm m_{ct}\sqrt{m_{ut}}  & \pm\sqrt{m_{ct}}(1-m_{uc}/2-m_{ct}/2) & \pm(1-m_{ct}/2) \end{array}\right),
\end{equation}}
\noindent
where $m_{ij}\equiv m_i/m_j,\;\; i<j;\;\; i,j=1,2,3=u,c,t$ respectively, and we have make $m_{ut}=0$ (but keeping $\sqrt{m_{ut}}\approx 10^{-3}\neq 0$).
From the former analysis we can evaluate the mixing matrix $V_{CKM}^{(f6)}=O_u^{(f6)}U_uU_d^\dagger O_d^{(f6)T}$, where $U_u$ and $U_d$ are as defined in (\ref{upph}). Explicitly, the elements of the CKM mixing matrix can be expressed as:
\begin{eqnarray}\label{vckm6}
(V_{CKM}^{(f6)})_{lm}&=&(O_u^{(f6)})_{l1}(O_d^{(f6)})_{m1}\\ \nonumber
&&+e^{i\phi_1}(O_u^{(f6)})_{l2}(O_d^{(f6)})_{m2}\\ \nonumber
&&+e^{i\phi_2}(O_u^{(f6)})_{l3}(O_d^{(f6)})_{m3},
\end{eqnarray}
where $\phi_1=(\alpha_u-\alpha_d)$ and $\phi_2=(\alpha_u+\beta_u-\alpha_d-\beta_d)$. Written in the previous form, $V_{CKM}^{(f6)}$ includes two different phases, $\phi_1$ and $\phi_2$, but since it is a well known fact that the SM mixing matrix can be parametrized with only one single phase, our analysis makes sense only for the following three different cases:
\begin{itemize}
 \item Case 1: $\phi_1\neq 0,\;\;\phi_2=0$.
 \item Case 2: $\phi_1 = 0,\;\;\phi_2\neq 0$.
 \item Case 3: $\phi_1=\phi_2\neq 0$.
\end{itemize}

The prediction for the Cabibbo angle from this six parallel texture zeros ``anz\"atze" can be extracted from the following analytic expression:
\begin{equation}\label{us6f}
V^{(f6)}_{us}=A(m_{ln})-e^{i\phi_1}B(m_{ln})-e^{i\phi_2}C(m_{ln}),
\end{equation}
with $l,n=u,c,t,d,s,b$, and 
\begin{equation}\label{ust6f}
\begin{split}
A&\approx\sqrt{m_{ds}{\frac{\Delta^+_{db}}{\Delta^+_{uc}\Delta^+_{ds}\Delta^+_{sb}(\Delta^-_{sb}+m_{db})}}} \\ 
B&\approx\sqrt{m_{uc}\frac{\Delta^-_{ct}\Delta^+_{db}}{\Delta^+_{uc}\Delta^+_{sb}\Delta^+_{ds}}}\\
C&\approx\sqrt{m_{ut}m_{sb}\frac{\Delta^-_{uc}\Delta^-_{sb}\Delta^-_{ds}}
{(\Delta^+_{uc}-m_{ct})\Delta^+_{ds}\Delta^+_{sb}(\Delta^-_{sb}+m_{db})}},
\end{split}
\end{equation}
where $\Delta^\pm_{ln}=1\pm m_{ln}$. Since the term proportional to $e^{i\phi_2}$ in (\ref{us6f}), is three or more orders of magnitude smaller than the other two terms, we can write for the Cabibbo angle, in a very crude approximation
\begin{equation}\label{us6fap}
V^{(f6)}_{us}\approx\sqrt{m_{ds}}-e^{i\phi_1}\sqrt{m_{uc}}\approx\sin\theta_{12},
\end{equation}
form advocated in some papers dealing with parallel six texture zeros~\cite{frxi}. Equation (\ref{us6fap}), or more appropriate (\ref{us6f}) and (\ref{ust6f}) can be used to determine the magnitude of $\phi_1$ by fitting $|V_{us}|$ with current data. As a matter of fact, for case 1, $\phi_1=1.536$ and using the central values for the quark mass values quoted in the appendix, we get $|V_{us}|\approx 0.226$ (case 2 for $\phi_1=0$ does not have solution and case 3 for $\phi_1=\phi_2=1.527$ gives also $|V_{us}|\approx 0.226$).

In a similar way we can find, in the context of this ``ans\"atze", the analytic expression for all the other entries of $V_{CKM}^{(f6)}$ as functions of the quark masses (and the CP violating phase); in particular for $V_{cb}$ we have:
\begin{equation}\label{cb6f}
V^{(f6)}_{cb}=A^\prime(m_{ij})-e^{i\phi_1}B^\prime(m_{ij})+e^{i\phi_2}C^\prime(m_{ij}),
\end{equation}
where
\begin{equation}\label{cbt6f}
\begin{split}
A^\prime&\approx m_{sb}\sqrt{m_{uc}m_{db}\frac{\Delta^-_{ds}}{\Delta^+_{uc}\Delta^+_{sb}\Delta^-_{db}(\Delta^-_{sb}+m_{db})}} \\ 
B^\prime&\approx\sqrt{m_{sb}\frac{\Delta^-_{ds}}{\Delta^+_{ct}\Delta^+_{uc}\Delta^+_{sb}\Delta^-_{db}}}\\
C^\prime&\approx\sqrt{m_{ct}\frac{\Delta^-_{uc}\Delta^-_{sb}\Delta^+_{db}}
{\Delta^+_{uc}\Delta^+_{ct}\Delta^+_{sb}\Delta^-_{db}(\Delta^-_{sb}+m_{db})}},
\end{split}
\end{equation}
where again $A^\prime$ is 3 or more orders of magnitude smaller than $B^\prime$ and $C^\prime$. So, in a very crude approximation we may write
\begin{equation}\label{cb6fap}
V^{(f6)}_{cb}\approx -e^{i\phi_1}\sqrt{m_{sb}}+e^{i\phi_2}\sqrt{m_{ct}}\approx\sin\theta_{23},
\end{equation}
where $\phi_i,\;\;i=1,2$ are now fix values obtained from $V_{us}$.

Plugging numbers in Eq.~(\ref{cbt6f}) we get $|V_{cb}|=0.1434$ for Case 1, and $|V_{cb}|=0.073$ for Case 3. So, the parallel six texture zeros ``anz\"atze" is ruled out because it can not explain the experimentally measured values for $|V_{us}|$ and $|V_{cb}|$ simultaneously.

As anticipated, and in accord with our way of counting parameters, for this ``ans\"atze" the 3 mixing angles are predicted as functions of the six quark masses and the CP violating phase.


\section{\label{sec:sec5}Non parallel six texture zeros}
In a similar way, let us study the following non parallel six texture zeros mass matrices, for which only the up sector is of the nearest neighbor interaction form, and the mixing angles $\theta_{13}$ and $\theta_{23}$ between the third family with the first two ones, came only from this quark sector:
\begin{equation}\label{textnpp}
M_u^{(np)} =\left(\begin{array}{ccc} 
0 & a_ue^{i\alpha_u} & 0 \\ 
a_ue^{-i\alpha_u} & 0 & b_ue^{i\beta_u}  \\
0 & b_ue^{-i\beta_u}  & c_u
\end{array}\right).
\end{equation}
\begin{equation}\label{textnp}
M_d^{(np)} =\left(\begin{array}{ccc} 
0 & a_de^{i\alpha_d} & 0 \\ 
a_de^{-i\alpha_d} & b_d & 0  \\
0 & 0  & c_d
\end{array}\right),
\end{equation}
(a parallel six texture zeros with $M_u$ similar to $M_d^{(np)}$ implies $\theta_{13}=\theta_{23}=0$ which is ruled out).

In a trivial way, the phases $\alpha_u$, and $\beta_u$ are removed with $U_u={\rm Diag}(1,e^{i\alpha_u},e^{i(\alpha_u+\beta_u)})$ as in (\ref{upph}), and $\alpha_d$ is removed with a transformation $U_d^\prime={\rm Diag}(1,e^{i\alpha_d},e^{i\alpha_d})$. For diagonalizing  $M_u^{(np)\prime}=U_uM_u^{(np)}U_u^\dagger$, the exact orthogonal matrix $O_u^{(f6)}$ can be used [or in its defect the approximate form given in (\ref{rot6ex})]. For diagonalizing the down quark sector $M_d^{(np)\prime}=U_d^{\prime}M_u^{(np)}U_d^{\prime\dagger}$, the $3\times 3$ matrix invariants allow us to write
\begin{equation}\label{cq6d}
\begin{split}
c_d&=m_b \\ 
b_d&=m_d-m_s<0\\ 
a_d&=\sqrt{m_dm_s},
\end{split}
\end{equation}
with the exact diagonalizing transformation $O_d^{(np)}$ given now by
\begin{equation}\label{rotnp}
O_d^{(np)}= \left(\begin{array}{ccc}
     \sqrt{\frac{m_s}{m_d+m_s}} &  \sqrt{\frac{m_d}{m_d+m_s}}  & 0 \\
   -\sqrt{\frac{m_d}{m_d+m_s}}   &  \sqrt{\frac{m_s}{m_d+m_s}}   & 0 \\
        0 & 0 & 1 
       \end{array}\right).
\end{equation}

The mixing matrix is now $V_{CKM}^{(np)}=O_u^{(f6)}U_uU_d^{\prime\dagger}O_d^{(np)T}$ with the complex entries coming from the diagonal matrix $U_uU_d^{\prime\dagger}={\rm Diag}(1,e^{i\phi_1},e^{i\phi_2^\prime})$ where $\phi_1=\alpha_u-\alpha_d$ and $\phi_2^\prime=(\alpha_u+\beta_u-\alpha_d)$, but due to the particular form of matrix (\ref{rotnp}), $\phi_2$ does not play any active role in the mixing matrix.

Calculated as before, the analytic expression for $V_{us}^{(np)}$ is now
\begin{equation}\label{vusnp}
V_{us}^{(np)}\approx-\sqrt{\frac{m_{ds}}{\Delta^+_{uc}\Delta^+_{ds}}}
+e^{i\phi_1}\sqrt{\frac{m_{uc}\Delta^-_{ct}}{\Delta^+_{ds}\Delta^+_{uc}}},
\end{equation}
which again reproduces the approximate form in (\ref{us6fap}). Taking $\phi_1=1.50$ in the former expression produces a value $|V_{us}^{(np)}|\approx 0.225$, in agreement with the experimental value.

In a similar way, analytic expressions for the other elements of the mixing matrix, in the context of this ``anz\"atze" can be evaluated. In particular we have
\begin{equation}\label{vcbnp}
V_{cb}^{(np)}\approx e^{i\phi_2}\sqrt{\frac{m_{ct}}{\Delta^+_{uc}\Delta^+_{ct}}}\sim\sqrt{m_{ct}},
\end{equation}
which is too large for the quark mass values quoted in the appendix. Any way, the random numerical analysis shows that, by plugging for the quark masses the values (in units of GeV's) $m_t=174,\;\; m_c=-0.320, \;\; m_u=0.002,\;\; m_b=2.89,\;\; m_s=-0.065$ and $m_d=0.0031$ in $M_u^{(np)}$ in (\ref{textnpp}) and in $M_d^{(np)}$ in (\ref{textnp}), with $\alpha_u=1.50 + \alpha_d$, and $\beta_u=\alpha_d-\alpha_u$ we obtain the following absolute values for the mixing matrix
\begin{equation}\label{manp}
 V_{CKM}^{(np)}=
\left(\begin{array}{ccc}
0.970 & 0.226 & 0.003 \\
0.226 & 0.973 & 0.042 \\
0.009 & 0.042 & 0.999
\end{array}\right),
\end{equation}
in good agreement with the experimental measured values presented above.

Also, for $\phi_1=1.50$ and $\phi_2^\prime=0$, and using the definitions for the 3 angles of the unitary triangle as in equations (\ref{ts}), (\ref{tc}) and (\ref{tb}), the following values of the B decays CP asymmetries, are obtained:
\[(\alpha,\;\beta,\;\gamma)_{th}^{(np)}=(89.02,\;\;\;20.05,\;\;\;71.2),\]
numbers which not only close the triangle as they should, but which are in good agreement with the observed experimental values.

Unfortunately, a close look at the former numerology shows that all the quark masses are in the experimentally allowed range, but the charm quark mass $m_c$ fails too short compared with the experimental accepted value.


\section{\label{sec:sec6}Five texture zeros}
From the former results, it seems obvious that a set of consistent $3\times 3$ quark mass matrices should emerge, just by modifying the previous non-parallel six texture zeros ``anz\"atze", by introducing an extra parameter in the up quark sector able to take care of the charm quark mass. With this in mind, let us use for the up quark sector the Hermitian two texture zeros mass matrix

\begin{equation}\label{text4}
M_u^{(4)} =\left(\begin{array}{ccc} 
0 & a_ue^{i\alpha_u} & 0 \\ 
a_ue^{-i\alpha_u}_q & d_u & b_ue^{i\beta_u}  \\
0 & b_ue^{-i\beta_u}  & c_u
\end{array}\right),
\end{equation}
with the down quark mass matrix given again  by (\ref{textnp}). As before, the two phases in (\ref{text4}) can be absorbed by working with the new quark fields $u_0^\prime,\; c_0^\prime$ and $t_0^\prime$ as defined in (\ref{upph}). Now there are seven real parameters to explain 6 masses and 3 mixing angles, so there must exist two physical prediction in the context of this ``ans\"atze".

The quark mass matrix in Eq. (\ref{text4}) has been analyzed in some detail in Refs.~\cite{verma}, let us review the main results: the invariants tr$[M_u^{(4)\prime}]$, det$[M_u^{(4)\prime}]$, and tr$[(M_u^{(4)\prime})^2]$ allow us to write the elements $a_u,\; b_u$ and $d_u$ of the mass matrix, in terms of the up quark masses and of the parameter $c_u$ as follows:
\begin{equation}\label{cq4}
\begin{split}
d_u&=m_u-m_c+m_t-c_u\\ 
a_u^2&=\frac{m_um_cm_t}{c_u}\\ 
b_u^2&=\frac{(c_u-m_u)(c_u+m_c)(m_t-c_u)}{c_u},
\end{split}
\end{equation}
and the exact diagonalizing transformation $O_u^{(4)}$ can be expressed as~\cite{verma}
{\small
\begin{equation}\label{rotf4}
\left(\begin{array}{ccc} 
\pm\sqrt{\frac{m_2m_3(c_u-m_1)}{(m_3-m_1)(m_2+m_1)c_u}} & \pm\sqrt{\frac{m_1m_3(c_u+m_2)}{(m_2+m_1)(m_3+m_2)c_u}} & \pm\sqrt{\frac{m_1m_2(m_3-c_u)}{(m_2+m_3)(m_3-m_1)c_u}} \\ 
\pm\sqrt{\frac{m_1(c_u-m_1)}{(m_3-m_1)(m_1+m_2)}} & \mp\sqrt{\frac{m_2(c_u+m_2)}{(m_2+m_3)(m_1+m_2)}} & \pm\sqrt{\frac{m_3(m_3-c_u)}{(m_2+m_3)(m_3-m_1)}}\\ 
\mp\sqrt{\frac{m_1(m_3-c_u)(m_2+c_u)}{(m_3-m_1)(m_1+m_2)c_u}} &
\pm\sqrt{\frac{m_2(m_3-c_u)(m_1-c_u)}{(m_2+m_1)(m_3+m_2)c_u}} & 
\pm\sqrt{\frac{m_3(c_u-m_1)(c_u+m_2)}{(m_2+m_3)(m_3-m_1)c_u}}\end{array}\right)^T,
\end{equation}
where notice that for $d_u=0$, the expressions in (\ref{cq4}) and (\ref{rotf4}) reduce to expressions (\ref{cq}) and (\ref{rotf6}) respectively. The use of (\ref{rotf4}), combined with (\ref{rotnp}) and (\ref{upph}), allow us to write the following analytic mixing matrix for $\phi_2=0$ [an irrelevant phase due to the particular form of the down quark mass matrix in (\ref{textnp})].

\[\left(
\begin{array}{ccc}
\frac{\sqrt{c_u-mu} \left(\sqrt{m_c m_s m_t}+e^{i\phi_1} \sqrt{c_u m_d m_u}\right)}
{\sqrt{c_u(m_d+m_s)(m_t-m_u) (m_c+m_u)}} 
& \frac{\sqrt{c_u-m_u} \left(-\sqrt{m_c m_d m_t}+e^{i\phi_1} \sqrt{c_um_sm_u}\right)}
{\sqrt{c_u(m_d+m_s)(m_t-m_u)(m_c+m_u),}}
& -\frac{\sqrt{(c_u + m_c) (-c_u+m_t) m_u}}{\sqrt{c_u (m_t-m_u) (m_c+m_u)}} \\
\frac{-e^{i\phi_1} \sqrt{c_u m_c m_d (c_u+m_c)}+\sqrt{(c_u+m_c) m_s m_t m_u}}
{\sqrt{c_u(m_d+m_s) (m_c+m_t) (m_c+m_u)}} 
& -\frac{e^{i\phi_1} \sqrt{c_u m_c m_s}+\sqrt{m_d m_t m_u}}{\sqrt{\frac{c_u (m_d+ m_s) (m_c+m_t) (mc+m_u)}{c_u+m_c}}} 
& \frac{\sqrt{m_c(-c_u+m_t)(c_u-m_u)}}{\sqrt{c_u(m_c+m_t)(m_c+m_u),}}  \\
\frac{\sqrt{-c_u+ m_t} \left(e^{i\phi_1} \sqrt{c_u m_d m_t}+\sqrt{m_c m_s m_u}\right)}
{\sqrt{c_u(m_d+m_s) (m_c+m_t) (m_t-m_u)}} 
& \frac{\sqrt{-c_u+m_t} \left(e^{i\phi_1} 
\sqrt{c_um_sm_t}-\sqrt{m_cm_dm_u}\right)}{\sqrt{c_u (m_d+m_s) (m_c+m_t) (m_t-m_u)}}
& \frac{\sqrt{(c_u+m_c)m_t (c_u-m_u)}}{\sqrt{c_u (m_c+m_t) (m_t-m_u)}}
\end{array}
\right).\]\\

From the former mixing matrix we can extract first 
\begin{equation}\label{vcb64}
V_{cb}=\frac{\sqrt{m_c(-c_u+m_t)(c_u-m_u)}}{\sqrt{c_u(m_c+m_t)(m_c+m_u),}}
\end{equation}
which we use to fix the free parameter $c_u$. Then we have

\begin{equation}\label{vus64}
V_{us}=\frac{\sqrt{c_u-m_u} \left(-\sqrt{m_c m_d m_t}+e^{i\phi_1} \sqrt{c_um_sm_u}\right)}
{\sqrt{c_u(m_d+m_s)(m_t-m_u)(m_c+m_u),}}
\end{equation}
which we use to fix the CP violating phase $\phi_1$. Now, using for the quark masses and for the $c_u$ parameter the values (given in units of GeV's):
$m_u=0.0024, \;\; m_c=-0.560, \;\; m_t=172, \;\; m_d=0.0029 \;\; m_s=-0.06  \;\; m_b=2.89$, and $c_u=171.721$, we obtain for a value $\phi_1=1.6$, the following numeric $3\times 3$ mixing matrix:

\[\left(
\begin{array}{ccc}
 0.97428 & 0.22532  & 0.00264 \\
 0.22517 & 0.97349  & 0.04020  \\
 0.00865 & 0.03934  & 0.99919
\end{array}
\right),\]
which is in quite good agreement with the experimental measured values. Finally, the three angles of the unitary triangle of the B decays CP asymmetries are calculated to be:
\[ (\alpha ,\; \beta ,\; \gamma)_{th}^{(5t)}= (90.79,\; 16.51, \; 72.68), \]
which not only close the triangle, but are such that $\alpha$ and $\gamma$ agree with the measured value at $1\sigma$ and $\beta$ at $2\sigma$.


\section{\label{sec:sec7}Conclusions}
In this note we have reported our finding of a simple pattern of quark mass matrices $M_u$ and $M_d$, which in the context of the SM are self-consistent and predict experimentally favored relations between quark masses and flavor mixing parameters. Our result points towards a five texture zeros quark mass matrices, for which the mixing angles $\theta_{13}$ and $\theta_{23}$ between the third family with the first two ones, came only from the up quark sector

Let us quote a few concluding remarks:

1. In the context of the SM or its extensions which have no flavor-changing right-handed currents, the quark mass matrices $M_u$ and $M_d$ can be taken to be Hermitian without loss of generality. Non Hermitian quark mass matrices are relevant only when physics beyond the SM is being considered.

2. Three texture zeros or less in the quark mass matrices of the SM can always be obtained in a trivial way by using weak basis transformations, and do not imply predictions between the elements of the flavor mixing matrix.

3. Four and five texture zeros imply one and two physical relationships respectively.

4. Six texture zeros imply 3 physical relationships, and allows the writing of the 3 mixing angles as functions of the six quarks masses and the CP violation phase.

5. More than six texture zeros are not possible because they imply either Det$M_q=0$ which is valid only in the chiral limit, or a degenerate quark mass spectrum, both situations incompatible with the real world.



The $2\sigma$ deviation of our calculated $\beta$ angle can imply one or several of the following possibilities:
\begin{itemize}
\item The measured $\beta$ value points toward physics beyond the SM.
\item The controversy about the precise measurement of $\beta_{exp}^{fit}$ has not been settled yet~\cite{sony}.
\item Our five texture zeros ``ans\"atze" does not work quite properly.
\end{itemize}

To conclude, let us write the five texture zeros we have found, using the following perturbation expansion parametrization for the mass matrices 

\begin{equation}\label{par6}
M_u =\frac{h_t}{2}\left(\begin{array}{ccc} 
0 & 2\lambda^6 & 0 \\ 
2\lambda^6 & -4\lambda^5 & 5\lambda^3  \\
0 & 5\lambda^3  & 2
\end{array}\right),
\end{equation}
and 
\begin{equation}\label{par4}
M_d =h_t\left(\begin{array}{ccc} 
0 & 1.3\lambda^7 & 0 \\ 
1.3\lambda^7 & -1.4\lambda^6 & 0  \\
0 & 0  & \lambda^3
\end{array}\right),
\end{equation}
for the values $h_t=171.72$ and $\lambda=0.25$.
\appendix
\section{Quark Masses}
\label{aA}
For the inputs used for carrying out the numerical calculations in the main text, we have adopted the following ranges of quark masses~\cite{Qexp} at the $M_Z$ energy scale (where the $V_{CKM}$ matrix elements in (\ref{maexp}) are measured)

\vspace{1cm}

\begin{tabular}{||l|l||}\hline
Up sector & Down sector \\ \hline
$m_t=171.7\pm 3.0$ GeV & $m_b=2.89\pm 0.09$ GeV\\
$m_c=0.619\pm 0.084$ GeV & $m_s=55^{+16}_{-15}$ MeV \\
$m_u=1.27^{+0.5}_{-0.42}$ MeV & $m_d=2.90^{+1.24}_{-1.19}$ MeV \\ \hline\hline
\end{tabular}

\vspace{0.9cm}

The light quark masses $m_u,\; m_d$ and $m_s$ can be further constrained using the mass ratios~\cite{leut}
\begin{equation}\label{mara}
m_u/m_d=0.553\pm 0.043\;,  \hspace{1cm} m_s/m_d=18.9\pm 0.8\;\;.
\end{equation}

Notice also that due to the experimental errors 
\[m_t\pm m_c\pm m_u\approx m_t.\]

{\it Note added in proof}. \\
After submission of this manuscript, we became aware of the existence of a paper by H. D. Kim, S. Raby and L. 
Schradin~\cite{kim} in which apparently, a different conclusion to ours was reached, with a $\sin(2\beta)$ value 
too small compared with the measured value $(3.5 \sigma$ off).

Two comments:\\
1. Kim {\it et al} use in their analysis quark masses at different scales, most of them pole values. We use all quark 
masses at the
$M_Z$ energy scale were we believe the $V_{CKM}$ matrix elements are measured. Although the mass quotients should not change
much, the error bars are quite different, and plugging in our numbers in their analytic results (Eq. 4.7 in~\cite{kim}) we 
find a $\beta$ value within a $1\sigma$ of the experimental measured value.\\
2. They use an approximation for $|V_{us}|$ (see Eq. 4.3 in~\cite{kim}). We use for $|V_{us}|$ an exact value 
(see Eq. [\ref{vus64} above] where $c_u$ is a new free parameter that must be fine tunned to the value $c_u=171.721$ GeV. 
in order to get agreement with the $V_{CKM}$ measured values [note that for $c_u\equiv m_t$ our Eq. (\ref{vus64}) reproduce 
Eq. (4.3) in~\cite{kim}].  


%
%
%
%

\begin{thebibliography}{}
\bibitem[1]{SM}
For an excellent compendium of the SM, see J. F. Donoghue, E, Golowich, and B. R. Holstein, {\it Dynamics of the Standard Model} (Cambridge University Press, Cambridge, U.K. 1992).
\bibitem[2]{wein}
S.Weinberg, Phys. Rev. D\textbf{5}, 1962 (1972); A.Zee, Phys. Lett. B\textbf{93}, 389 (1980); A.Zee, Nucl. Phys. B\textbf{264}, 99 (1986); K.S.Babu, Phys. Lett. B\textbf{203}, 132 (1988).
\bibitem[3]{Dis}
F. Wilczek and A. Zee, Phys. Lett. B\textbf{70}, 418 (1977); S. Pakvasa and H. Sugawara, Phys. Lett. B\textbf{73}, 61 (1978); Y. Yamanaka, H. Sugawara, and S. Pakvasa, Phys. Rev. D\textbf{25}, 1895 (1982); K.S. Babu and X.G. He, Phys. Rev. D\textbf{36}, 3484 (1987)
\bibitem[4]{Con}
F. Wilczek and A. Zee, Phys. Rev. Lett. \textbf{42}, 421 (1979); A. Davidson, M. Koca, and K.C. Wali, Phys. Rev. D\textbf{43}, 92 (1979).
\bibitem[5]{Fro}
C.D. Froggatt and H.B. Nielsen, Nucl. Phys. B\textbf{147}, 277 (1979).
\bibitem[6]{Fri} 
H. Fritzsch, Phys.Lett.B\textbf{73}, 317 (1978); Nucl. Phys. B\textbf{155}, 182, (1979). For a review with extensive references, see: H.Fritzsch and Z.Z.Xing, Prog. Part. Nucl. Phys. \textbf{45}, 1 (2000).
\bibitem[7]{Ram} 
P. Ramond, R. G. Roberts, and G. G. Ross, Nucl. Phys., B\textbf{406}, 19 (1993); M.Leurer, Y. Nir and N. Seiberg, Nucl. Phys. B\textbf{398}, 319 (1993): L.Iba\~nez and G.G.Ross, Phys. Lett.B\textbf{332}, 100 (1994).
\bibitem[8]{koma}
M.Kobayashi and T.Maskawa, Prog. Theor. Phys. \textbf{49}, 652 (1973); L.Maiani, Phys. Lett. B\textbf{62}, 183 (1976)
\bibitem[9]{pdg} K.Nakamura \textit{et al.} (Particle Data Group), J. Phys., G\textbf{37}, 075021 (2010); http://pdg.lbl.gov/.
\bibitem[10]{Bra1}
G.C.Branco, L.Lavoura, and F.Mota, Phys.Rev. D\textbf{39}, 3443 (1989); H.Fritzsch and Z.Z.Xing, Nucl. Phys. B\textbf{556}, 49 (1999);
G.C.Branco, D. Emmanuel-Costa and R. Gonz\'alez Felipe, Phys. Lett. B\textbf{477}, 147 (2000).
\bibitem[11]{nir}
C.Dib, I.Dunietz, F.J.Gilman, and Y.Nir, Phys. Rev. D\textbf{41}, 1522 (1990).
\bibitem[12]{ckmf}
CKMfitter Group (J.Charles \textit{et al.}), Eur.Phys.J. C \textbf{41},1 (2005). Updated results and plots available at: http://ckmfitter.in2p3.fr/.
\bibitem[13]{babar}
P.F.Harrison and H.R.Quinn, editors [BABAR Collab.], ``The BABAR Physics book": physics at an asymmetric B factory", SLAC-R-0504.
\bibitem[14]{sony}
Y. Nir, Nucl. Phys. Proc. Suppl. \textbf{117}, 111 (2003); E. Lunghi and A. Soni, Phys. Rev Lett. \textbf{104}, 251802 (2010).
\bibitem[15]{frxi}
H. Fritzsch and Z.Z.Xing, Phys. Lett B\textbf{555}, 63 (2003).
\bibitem[16]{verma}
H. Fritzsch and Z.Z. Xing, Phys.Lett. B\textbf{353}, 114 (1995); Phys. Lett. B\textbf{555}, 63 (2003); 
Z.Z.Xing and H.Zhang, J. Phys. G\textbf{30}, 129 (2004); R.Verma \textit{et al}, J. Phys. G: Nucl. Part. Phys. \textbf{37} 075020 (2010).
\bibitem[17]{Qexp} 
For updated Values of Running Quark and Lepton Masses see: Zhi-zhong Xing, He Zhang, Shun Zhou, Phys.Rev.D77:113016,2008 [arXiv:0712.1419].
\bibitem[18]{leut}
H. Leutwyler, Phys. Lett. B\textbf{378}, 313 (1996).
\bibitem[19]{kim}
H. D. Kim, S. Raby and L. Schradin, Phys. Rev. D \textbf{69}, 092002 (2004).
\end{thebibliography}
%

\end{document}